\def\papertitle{Towards multi-instrument drum transcription}

\documentclass[twoside,a4paper]{article}
\usepackage{dafx_18}
\usepackage{amsmath,amssymb,amsfonts,amsthm}
\usepackage{euscript}
\usepackage[T1]{fontenc}
\usepackage{ifpdf}

\usepackage[english]{babel}
\usepackage{caption}
\usepackage{subfig} 
\usepackage{color}

\usepackage{multirow} 
\usepackage{arydshln} 
\usepackage{todonotes}
\usepackage{pgf}

\ninept

\usepackage{times}

\newif\ifpdf
\ifx\pdfoutput\relax
\else
   \ifcase\pdfoutput
      \pdffalse
   \else
      \pdftrue
\fi

\ifpdf 
\else 
  \usepackage[dvips]{epsfig,graphicx}
\fi

\title{\papertitle}

\affiliation
{Richard Vogl\textsuperscript{1,2} \hspace{3.5cm} Gerhard Widmer\textsuperscript{2} \hspace{3.5cm} Peter Knees\textsuperscript{1}}
{	\vspace{-0.8em}{\small \tt{richard.vogl@tuwien.ac.at} \hspace{1.0cm} \tt{gerhard.widmer@jku.at} \hspace{1.0cm} \tt{peter.knees@tuwien.ac.at} } \\
	\vspace{0.8em}
	\begin{minipage}{.45\textwidth}
		\centering
		\textsuperscript{1} Faculty of Informatics \\ TU Wien \\ Vienna, Austria \\ 
	\end{minipage}%
	\begin{minipage}{.45\textwidth}
		\centering
		\textsuperscript{2} Institute of Computational Perception \\ Johannes Kepler University \\ Linz, Austria\\
	\end{minipage}
}

\begin{document}
\ifpdf 
  \DeclareGraphicsExtensions{.png,.jpg,.pdf}
\else  
  \DeclareGraphicsExtensions{.eps}
\fi

\newcommand{\smllab}{3}
\newcommand{\midlab}{8}
\newcommand{\lrglab}{18}

\newcommand{\cnn}{CNN}
\newcommand{\crnn}{CRNN}

\def\ppfun{f_a}

\def\enst{\emph{ENST}}
\def\smt{\emph{SMT}}
\def\rbma{\emph{RBMA13}}
\def\mdb{\emph{MDB}}
\def\all{\emph{all}}
\def\midi{\emph{MIDI}}
\def\miditen{\emph{MIDI 10\%}}
\def\midione{\emph{MIDI 1\%}}
\def\midibal{\emph{MIDI bal.}}
\def\ptmidi{\emph{pt MIDI}}
\def\ptmidibal{\emph{pt MIDI bal.}}

\let\pgfimageWithoutPath\pgfimage 
\renewcommand{\pgfimage}[2][]{\pgfimageWithoutPath[#1]{#2}}

\maketitle

\begin{abstract}
\looseness -1
Automatic drum transcription, a subtask of the more general automatic music transcription, deals with extracting drum instrument note onsets from an audio source.
Recently, progress in transcription performance has been made using non-negative matrix factorization as well as deep learning methods.
However, these works primarily focus on transcribing three drum instruments only: snare drum, bass drum, and hi-hat.
Yet, for many applications, the ability to transcribe more drum instruments which make up standard drum kits used in western popular music would be desirable.
In this work, convolutional and convolutional recurrent neural networks are trained to transcribe a wider range of drum instruments.
First, the shortcomings of publicly available datasets in this context are discussed.
To overcome these limitations, a larger synthetic dataset is introduced.
Then, methods to train models using the new dataset focusing on generalization to real world data are investigated.
Finally, the trained models are evaluated on publicly available datasets and results are discussed.
The contributions of this work comprise:
\emph{(i.)}~a large-scale synthetic dataset for drum transcription, \emph{(ii.)}~first steps towards an automatic drum transcription system that supports a larger range of instruments by evaluating and discussing training setups and the impact of datasets in this context, and \emph{(iii.)}~a publicly available set of trained models for drum transcription.
Additional materials are available at \texttt{http://ifs.tuwien.ac.at/\textasciitilde vogl/dafx2018}.
\end{abstract}

\section{Introduction}
\label{sec:introduction}

Automatic drum transcription (ADT) focuses on extracting a symbolic notation for the onsets of drum instruments from an audio source.
As a subtask of automatic music transcription, ADT has a wide variety of applications, both in an academic as well as in a commercial context.
While state-of-the-art approaches achieve reasonable performance on publicly available datasets, there are still several open problems for this task.
In prior work \cite{VoglDWK17_DrumTransCRNN_ISMIR} we identify additional information---such as bar boundaries, local tempo, or dynamics---required for a complete transcript
and propose a system trained to detect beats alongside drums.
While this adds some of the missing information, further work in this direction is still required.

Another major shortcoming of current ap\-proaches is the limitation to only three drum instruments. 
The focus on snare drum (SD), bass drum (BD), and hi-hat (HH) is motivated by the facts that these are the instruments \emph{(i.)} most commonly used and thus with the highest number of onsets in the publicly available datasets; and \emph{(ii.)} which often define the main rhythmical theme. 
Nevertheless, for many applications it is desirable to be able to transcribe a wider variety of the drum instruments which are part of a standard drum kit in western popular music, e.g., for extracting full transcripts for further processing in music production or educational scenarios.
One of the main issues with building and evaluating such a system is the relative underrepresentation of these classes in available datasets (see section~\ref{sec:relwork}).

In this work we focus on increasing the number of instruments to be transcribed.
More precisely, instead of three instrument classes, we aim at transcribing drums at a finer level of granularity as well as additional types of drums, leading to classification schemas consisting of eight and 18 different instruments (see table~\ref{tab:labels}).
In order to make training for a large number of instruments feasible, we opt for a single model to simultaneously transcribe all instruments of interest, based on convolutional and convolutional recurrent neural networks.
Especially in the case of deep learning, a considerable amount of processing power is needed to train the models.
Although other approaches train separate models for each instrument in the three-instrument-scenario~\cite{southall2016automatic,SouthallSH17_DrumTransCNN_ISMIR}, for 18 instruments it is more feasible to train a single model in a multi-task fashion (cf.~\cite{caruana1998multitask}).
To account for the need of large volumes of data in order to train the chosen network architectures, a large synthetic dataset is introduced, consisting of 4197 tracks and an overall duration of about 259h.

The remainder of this paper is organized as follows.
In section~\ref{sec:relwork} we discuss related work, followed by a description of our proposed method in section~\ref{sec:method}.
Section~\ref{sec:datasets} provides a review of existing datasets used for evaluation, as well as a description of the new, large synthetic dataset.
Sections~\ref{sec:experiments} and \ref{sec:results} describe the conducted experiments and discuss the results, respectively.
Finally, we draw conclusions in section~\ref{sec:conclusion}.

\begin{table}[t]
        \caption{Classes used in the different drum instrument classification systems. Labels map to General MIDI drum instruments: e.g. bass drum: 35, 36; side stick: 37; etc. The mapping is available on the accompanying website.
        }
    \centering
        \begin{tabular}{| c | c | c | l |}
            \hline
            \multicolumn{3}{|c|}{number of classes} & \multirow{2}{*}{instrument name} \\
            \smllab & \midlab & \lrglab &  \\
            \hline
            \hline
            BD	&	BD &	BD & bass drum \\
            \hdashline
            SD	&	SD	&	SD & snare drum \\
            \hdashline
            &		     &  SS & side stick \\
            &		     &  CLP & hand clap \\
            \cdashline{2-4}
            &\multirow{3}{*}{TT}&  HT & high tom\\ 
            &		     &  MT & mid tom \\  
            &		     &  LT & low tom \\  
            \hdashline	  
            \multirow{3}{*}{HH}&\multirow{3}{*}{HH}&  CHH & closed hi-hat\\  
            &		     &  PHH & pedal hi-hat \\  
            &		     &  OHH & open hi-hat \\  
            \hdashline
            &		     &  TB & tambourine \\  
            \cdashline{2-4}	  
            &	 RD    &  RD & ride cymbal \\  
            \cdashline{2-4}
            &\multirow{2}{*}{BE}&  RB & ride bell \\  
            &		     &  CB & cowbell \\ 
            \cdashline{2-4}
            &\multirow{2}{*}{CY}&  CRC & crash cymbal \\  
            &		     &  SPC & splash cymbal \\  
            &		     &  CHC & Chinese cymbal \\ 
            \cdashline{2-4}
            &	 CL    &  CL & clave/sticks \\ 
            
            \hline
        \end{tabular}
    \label{tab:labels}
\end{table}

\vspace{-1mm}
\section{Related Work}\label{sec:relwork}

There has been a considerable amount of work published on ADT in recent years, e.g.,~\cite{gillet2004automatic,miron2013open,yoshii2007drum,paulus2009drum,Wu_2017_drumTransUnlabeledData_ismir}. 
In the past, different combinations of signal processing and information retrieval techniques haven been applied to ADT. 
For example: onset detection in combination with \emph{(i.)} bandpass filtering \cite{tzanetakis2005subband,KaliakatsosFVK12_DrumTrans_AudioMostly}, 
and \emph{(ii.)} instrument classification \cite{gillet2004automatic, miron2013open,yoshii2007drum};
as well as probabilistic models \cite{paulus2009drum, gillet2007supervised}.
Another group of methods focus on extracting an onset-pseudo-probability function (activation function) for each instrument under observation.
These methods utilize source separation techniques like Independent Subspace Analysis (ISA) \cite{FitzgeraldLC_DrumTransISA_DAFX}, Prior Subspace Analysis (PSA) \cite{SpichZST10_DrumTransPSA_DAFX}, 
and Non-Negative Independent Component Analysis (NNICA) \cite{dittmar2004further}.
More recently, these approaches have been further developed using Non-Negative Matrix Factorization (NMF) variants as well as deep learning \cite{VoglDWK17_DrumTransCRNN_ISMIR, SouthallSH17_DrumTransCNN_ISMIR, wu2015drum, dittmar2014real}.

The work of Wu et al.~\cite{wu2018survey} provides a comprehensive overview of the publications for this task, and additionally performs in-depth evaluation of current state-of-the-art methods.
Due to the large number of works and given the space limitations, in the remainder of this section, we will focus on work that is directly relevant with respect to the current state of the art and methods focusing on more than three drum instrument classes.

As mentioned, the state of the art for this task is currently defined by end-to-end activation function based methods.
In this context, end-to-end implies using only one processing step to extract the activation function for each instrument under observation from a digital representation of the audio signal (usually spectrogram representations). 
Activation functions can be interpreted as probability estimates for a certain instrument onset at each point in time.
To obtain the positions of the most probable instrument onsets, simple peak picking \cite{vogl2016recurrent, vogl2017transcription, VoglDWK17_DrumTransCRNN_ISMIR, SouthallSH17_DrumTransCNN_ISMIR, southall2016automatic, wu2015drum, dittmar2004further} or a language-model-style decision process like dynamic Bayesian networks \cite{bock2016joint} can be used.
These methods can be further divided into NMF based and deep neural network (DNN) based approaches.

Wu et al. \cite{wu2015drum} introduce partially fixed NMF (PFNMF) and further modifications to extract the drum instrument onset times from an audio signal.
Dittmar et al. \cite{dittmar2014real} use another modification of NMF, namely semi adaptive NMF (SANMF) to transcribe drum solo tracks in real time, while requiring samples of the individual drum instruments for training.
More recently, recurrent neural networks (RNNs) have successfully been used to extract the activation functions for drum instruments \cite{vogl2016recurrent, vogl2017transcription, southall2016automatic}. 
It has also been shown that convolutional (CNNs) \cite{VoglDWK17_DrumTransCRNN_ISMIR, SouthallSH17_DrumTransCNN_ISMIR} and convolutional recurrent neural networks (CRNNs) \cite{VoglDWK17_DrumTransCRNN_ISMIR} have the potential to even surpass the performance of RNNs.

The majority of works on ADT, especially the more recent ones, focus solely on transcribing three drum instrument (SD, BD, HH) \cite{Wu_2017_drumTransUnlabeledData_ismir, vogl2016recurrent, vogl2017transcription, VoglDWK17_DrumTransCRNN_ISMIR, southall2016automatic, SouthallSH17_DrumTransCNN_ISMIR, wu2015drum, paulus2009drum, dittmar2014real, yoshii2007drum, paulus2009drum}.
In some works multiple drum instruments are grouped into categories for transcription \cite{gillet2004automatic} and efforts have been made to classify special drum playing techniques within instrument groups~\cite{WuLerch16_DrumTechniqueNMF_ISMIR}.
However, only little work exists which approach the problem of transcribing more than three individual drum instruments \cite{dittmar2004further}, furthermore, such a system has---to our knowledge---never been evaluated on currently available public drum transcription datasets.

In \cite{miron2013open}, a set of MIDI drum loops rendered with different drum samples are used to create synthetic data in the context of ADT. 
Using synthetic data was a necessity in the early years of music information retrieval (MIR), but due to the continuous efforts of creating datasets, this has declined in recent years.
However, machine learning methods like deep learning, often requirer large amounts of data, and manual annotation in large volumes is unfeasible for many MIR tasks.
In other fields like speech recognition or image processing, creating annotations is easier, and large amounts of data are commonly available.
Using data augmentation can, to a certain degree, be used to overcome lack of data, as has been demonstrated in the context of ADT~\cite{vogl2017transcription}.
In~\cite{salamon2017analysis}  an approach to resynthesizes solo tracks using automatically annotated f0 trajectories, to create perfect annotations, is introduced.
This approach could be applicable for ADT, once a satisfactory model for the full range of drum instruments is available.
At the moment such annotations would be limited to the three drum instrument classes used in state-of-the-art methods.

\begin{figure}
    \centerline{
        \includegraphics[width=\columnwidth]{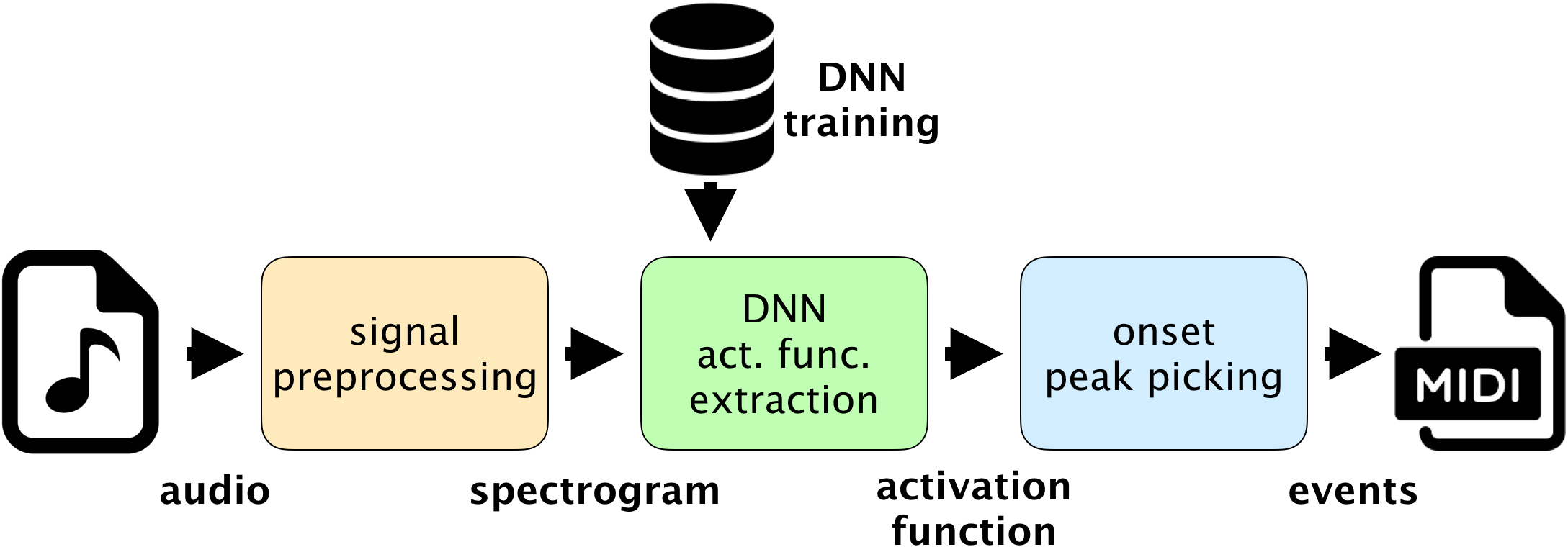}}
    \caption{Overview of implemented ADT system using DNNs. }
    \label{fig:overview}
\end{figure}
\vspace{-1mm}
\section{Method}\label{sec:method}

In this work, we use an approach similar to the ones introduced in \cite{southall2016automatic} and 
\cite{vogl2016recurrent}, for drum transcription. 
As mentioned in the introduction, a single model trained in a multi-task fashion will be used.
Creating individual models for each instrument is an option \cite{southall2016automatic, SouthallSH17_DrumTransCNN_ISMIR}, however, in the context of this work it has two downsides:
First, training time will scale linearly with the amount of models, which is problematic when increasing the number of instruments under observation.
Second, training multi-task models in the context of ADT can improve the performance~\cite{VoglDWK17_DrumTransCRNN_ISMIR}.
Other state-of-the-art methods based on NMF~\cite{wu2015drum, dittmar2014real} are less suitable for a multi-task approach, since the performance of NMF methods is prone to degrade for basis matrices with higher rank.

Thus, the method proposed in \cite{VoglDWK17_DrumTransCRNN_ISMIR} seems most promising for the goal of this work.
We will only use CNNs and CRNNs, since simple RNNs do not have any advantage in this context.
The implemented ADT system consists of three stages: a signal preprocessing stage, a DNN activation function extraction stage, and a peak picking post processing stage, identifying the note onset.
The system overview is visualized in figure~\ref{fig:overview}, and the single stages will be discussed in detail in the following subsections.

\subsection{Preprocessing}\label{sec:preprocessing}

During signal preprocessing, a logarithmic magnitude spectrogram is calculated using a window size of 2048 samples (@44.1kHz input audio frame rate) and choosing 441 samples as hop size for a 100Hz target frame rate of the spectrogram.
The frequency bins are transformed to a logarithmic scale using triangular filters in a range from 20 to 20,000 Hz, using 12 frequency bins per octave. 
Finally, the positive first-order-differential over time of this spectrogram is calculated and stacked on top of the original spectrogram.
The resulting feature vectors have a length of 168 values (2x84 frequency bins).
\vspace{-1mm}
\subsection{Activation Function Extraction}\label{sec:actfunc}
\begin{figure}[t]
    \centerline{
        \includegraphics[width=\columnwidth]{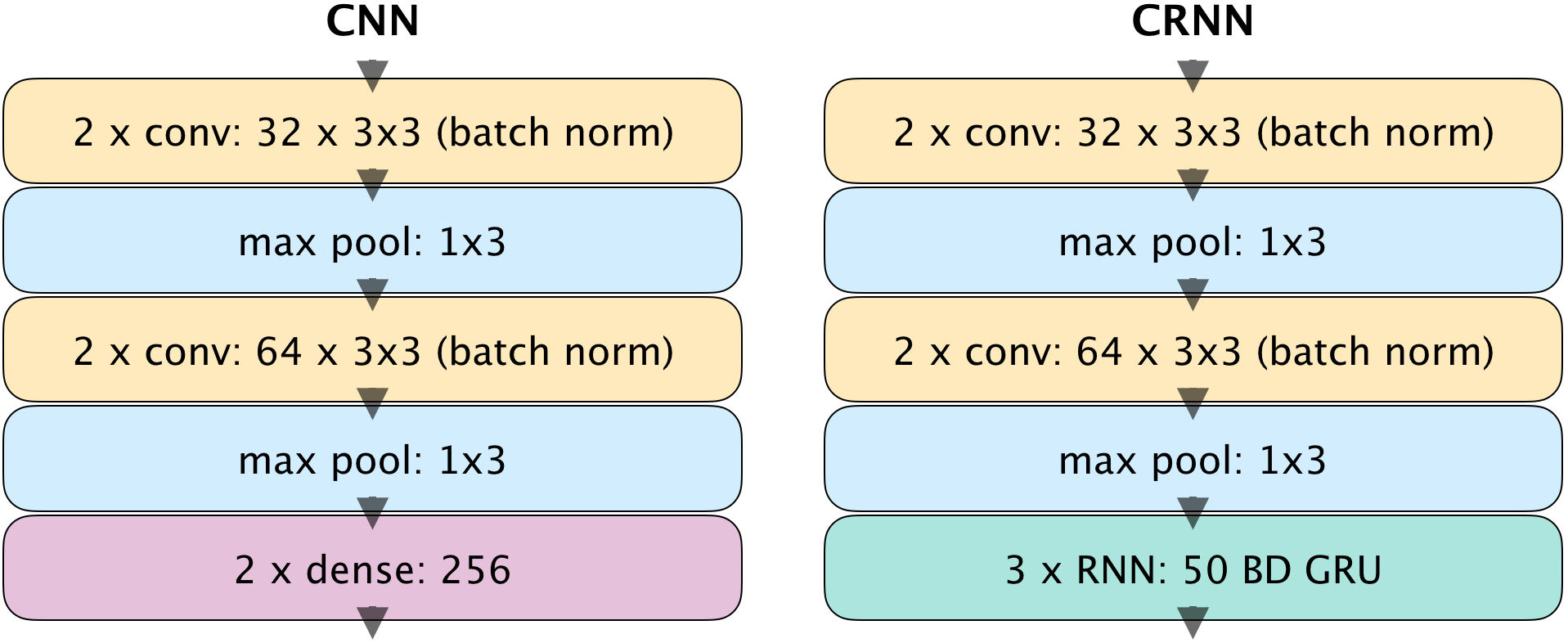}}
    \caption{Architecture comparison between the CNN and CRNN used for activation function extraction. }
    \label{fig:architectures}
\end{figure}

The activation function extraction stage is realized using one of two different DNNs architectures.
Figure~\ref{fig:architectures} visualizes and compares the two implemented architectures.
The convolutional parts are equivalent for both architectures, however, the dense output layers are different:
while for the CNN two normal dense layers are used (ReLUs), in case of the CRNN two bidirectional RNN layers consisting of gated recurrent units (GRUs) \cite{cho2014gru} are used. 
As already noted in \cite{VoglDWK17_DrumTransCRNN_ISMIR}, GRUs exhibit similar capabilities as LSTMs \cite{hochreiter1997lstm}, while being more easy to train.

The combination of convolutional layers which focus on local spectral features, and recurrent layers which model mid- and long-term relationships, has been found to be one of the best performing models for ADT \cite{VoglDWK17_DrumTransCRNN_ISMIR}.

\subsection{Peak Picking}\label{sec:peackpicking}

To identify the drum instrument onsets, a standard peak picking method introduced for onset detection in~\cite{boeck2013maximum} is used. 
A peak at point $n$ in the activation function $\ppfun(n)$ must be the maximum value within a window of size $m+1$ (i.e.: $\ppfun(n) = max (\ppfun(n-m), \cdots, \ppfun(n))$), and exceeding the mean value plus a threshold $\delta$ within a window of size $a+1$ (i.e.: $\ppfun(n) \geq mean(\ppfun(n-a), \cdots, \ppfun(n))+\delta$). 
Additionally, a peak must have at least a distance of $w+1$ to the last detected peak $n_{lp}$ (i.e.: $n-n_{lp} > w,$). 
The parameters for peak picking are the same as used in \cite{VoglDWK17_DrumTransCRNN_ISMIR}: $m = a = w = 2$.
The best threshold for peak picking is determined on the validation set. 
As observed in~\cite{SouthallSH17_DrumTransCNN_ISMIR,vogl2017transcription,VoglDWK17_DrumTransCRNN_ISMIR}, appropriately trained DNNs produce spiky activation functions, therefore, low thresholds ($0.1-0.2$) give best results.

\subsection{Training and Evaluation}
\looseness -1

Training of the models is performed using \emph{Adam} optimization~\cite{kingma2014adam} with mini-batches of size 100 and 8 for the CNNs and CRNNs respectively.
The training instances for the CNN have a spectral context of 25 samples.
In case of the CRNN, the training sequences consist of 400 instances with a spectral context of 13 samples.
The DNNs are trained using a fixed learning rate ($l_r=0.001$) with additional refinement if no improvement on the validation set is achieved for 10 epochs. 
During refinement the learning rate is reduced ($l_r=l_r \cdot 0.2$) and training continues using the parameters of the best performing model so far.

A three-fold cross-validation strategy is employed, using two splits during training, while 15\% of the training data is separated and used for validation after each epoch (0.5\% in case of the large datasets, to reduce validation time).
Testing is done on the third, during training unseen, split.
Whenever available, drum solo versions of the tracks are used as additional training material, but not for testing/evaluation.
The solo versions are always put into the same splits as their mixed counterparts, to counter overfitting.
This setup is consistently used through all experiments, whenever datasets are mixed or cross-validated, corresponding splits are used.

For audio preprocessing, peak picking, and calculation of evaluation metrics, the madmom\footnote{https://github.com/CPJKU/madmom} python framework was used.
DNN training was performed using Theano\footnote{https://github.com/Theano/Theano} and Lasagne\footnote{https://github.com/Lasagne/Lasagne}.
For a more details on C(R)NN training and a comparison of their working principles in the context of ADT, we kindly refer the reader to our previous work \cite{VoglDWK17_DrumTransCRNN_ISMIR} due to space limitations and a different focus of this work. 

\vspace{-3mm}
\section{Datasets}\label{sec:datasets}

\begin{figure}
    \centerline{
        \input{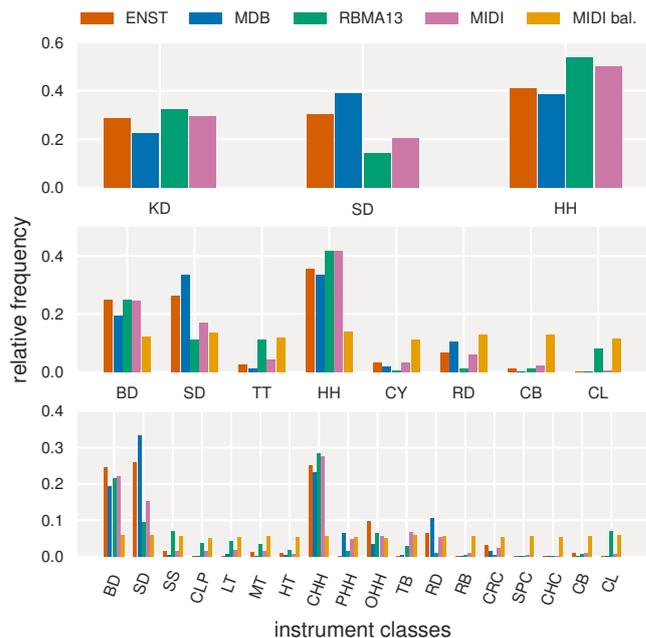}}
    \caption{Label distributions of the different datasets used in this work.}
    \label{fig:labeldist}
\end{figure}

There are a number of publicly available datasets for ADT with varying size, degree of detail, and number of classes regarding the drum instrument annotations. 
As noted in the introduction, current state-of-the-art approaches limit the instruments under observation to the three most common ones (SD, BD, HH).
This is done by ignoring other instruments like tom-toms and cymbals, as well as grouping different play styles like closed, opened, and pedal hi-hat strokes.
In order to investigate ways of generating a model which is capable to transcribe more than these three instruments, two classification systems, i.e., a medium and a large one, for drum instruments of a standard drum kit are defined.
Table~\ref{tab:labels} shows the two sets of classes, which contain eight and 18 labels respectively, alongside with the classic three-class set used in state-of-the-art works and the mapping used between these classes.

In the following we discuss publicly available ADT datasets and their limitations, leading to the description of the large volume synthetic dataset introduced for training of our models. 

\vspace{-1mm}
\subsection{ENST Drums (\enst)}
The ENST Drums\footnote{\texttt{http://perso.telecom-paristech.fr/\textasciitilde grichard/ENST-drums/}} dataset published by Gillet and Richard \cite{gillet2006enst} in 2005, is commonly used in ADT evaluations.
The freely available part of the dataset consists of single track audio recordings and mixes, performed by three drummers on different drum kits.
It contains recordings of single strokes for each instrument, short sequences of drum patterns, as well as drum tracks with additional accompaniment (\emph{minus-one} tracks).
The annotations contain labels for 20 different instrument classes.

For evaluation, the \emph{wet mixes} (contain standard post-processing like compression and equalizing) of the \emph{minus-one tracks} were used.
They make up 64 tracks of 61s average duration and a total duration of 1h.
The rest of the dataset (single strokes, patterns) was used as additional training data.

\begin{table}
	\caption{F-measure (\emph{mean}/\emph{sum}) results of implemented ADT methods on public datasets for different class systems. The first line indicates state-of-the-art F-measure results in previous work using CNN and CRNN ADT systems in a three-class scenario.}
	\centering
	\begin{tabular}{| r |  l | c | c | c | }
		\hline
		CL	&  model  & \enst & \mdb & \rbma \\  
		\hline
		\hline
		\emph{3} & \emph{SotA} \cite{VoglDWK17_DrumTransCRNN_ISMIR} & \hspace{1em}\emph{--- / 0.78}	& \emph{--- / ---}	& \hspace{1em}\emph{--- / 0.67}	\\    
		\hdashline
		\multirow{2}{*}{\smllab} & \cnn     & 0.75 / 0.77  & 0.65 / 0.72  & 0.53 / 0.63 \\
		                         & \crnn    & 0.74 / 0.76  & 0.64 / 0.70  & 0.55 / 0.64 \\
		\hdashline
		\multirow{2}{*}{\midlab} & \cnn 	& 0.59 / 0.63  & 0.68 / 0.65  & 0.55 / 0.44 \\
		                         & \crnn 	& 0.65 / 0.70  & 0.68 / 0.63  & 0.55 / 0.50 \\
		\hdashline
		\multirow{2}{*}{\lrglab} & \cnn     & 0.69 / 0.49  & 0.76 / 0.47  & 0.62 / 0.31 \\
		                         & \crnn    & 0.75 / 0.67  & 0.77 / 0.55  & 0.64 / 0.39 \\
		\hline
	\end{tabular}
	\label{tab:results_classic}
\end{table}
\vspace{-1mm}
\subsection{MDB-Drums (\mdb)}
The MDB-Drums dataset\footnote{\texttt{https://github.com/CarlSouthall/MDBDrums}} was published in \cite{southall2017mdb} and provides drum annotations for 23 tracks of the Medley DB dataset\footnote{\texttt{http://medleydb.weebly.com/}} \cite{bittner2014medleydb}.
The tracks are available as drum solo tracks with additional accompaniment.
Again, only the full mixes are used for evaluation, while the drum solo tracks are used as additional training data.
There are two levels of drum instrument annotations, the second providing multiple drum instruments and additional drum playing technique details in 21 classes.
Tracks have an average duration of 54 seconds and the total duration is 20m 42s.
\vspace{-1mm}
\subsection{RBMA13 (\rbma)}
The RBMA13 datasets\footnote{\texttt{http://ifs.tuwien.ac.at/\textasciitilde vogl/datasets/}} was published alongside \cite{VoglDWK17_DrumTransCRNN_ISMIR}.
It consists of 30 tracks of the freely available 2013 Red Bull Music Academy Various Assets sampler.\footnote{\texttt{https://rbma.bandcamp.com/album/}} 
The tracks' genres and drum sounds of this set are more diverse compared to the previous sets, making it a particularly difficult set.
It provides annotations for 23 drum instruments as well as beat and downbeats.
Tracks in this set have an average duration of 3m 50s and a total of 1h 43m.
\vspace{-1mm}
\subsection{Limitations of current datasets}
A major problem of publicly available ADT datasets in the context of deep learning is the volume of data.
To be able to train DNNs efficiently, usually large amounts of diverse data are used (e.g. in speech and image processing).
One way to counter the lack of data is to use data augmentation (as done in \cite{vogl2017transcription} for ADT).
However, data augmentation is only helpful to a certain degree, depending on the applicable augmentation methods and the diversity of the original data.

Given the nature of drum rhythms found in western popular music, another issue of ADT datasets is the uneven distribution of onsets between instrument classes.
In case of the available datasets, this imbalance can be observed in figure~\ref{fig:labeldist}.
While it is advantageous for the model to adapt to this bias, in terms of overall performance, this often results in the trained models to never predict onsets for sparse classes. 
This is due to the number of potential false negatives being negligible, compared to the amount of false positives produced in the early stages of training.
To counter a related effect on slightly imbalanced classes (BD, SD, HH in the three-class scenario), a weighting of the loss functions for the different classes can be helpful \cite{vogl2017transcription}.
Nevertheless, a loss function weighting cannot compensate for the problem in the case of very sparse classes.

Since manual annotation for ADT is a very resource intensive task, a feasible approach to tackle these problems is to create a synthetic dataset using the combination of symbolic tracks, e.g. MIDI tracks, drum synthesizers and/or sampler software.

\begin{table}
	\caption{F-measure results (\emph{mean}/\emph{sum}) of the implemented networks on synthetic datasets.}
	\centering
	\begin{tabular}{| l |  l | c | c | c | }
		\hline
		CL	&  model  & \midi & \midione & \midibal \\
		\hline
		\hline
		\multirow{2}{*}{\smllab} & \cnn    & 0.74 / {\bf 0.84}  & 0.70 / 0.79  & --- / --- \\
		                         & \crnn   & 0.74 / {\bf 0.84}  & 0.68 / 0.77  & --- / --- \\ 
		\hdashline
		\multirow{2}{*}{\midlab} & \cnn    & 0.64 / 0.63        & 0.63 / 0.69  & 0.54 / 0.58 \\ 
							     & \crnn   & 0.74 / {\bf 0.82}  & 0.69 / 0.73  & 0.58 / 0.70 \\ 
		\hdashline
         \multirow{2}{*}{\lrglab} & \cnn 	& 0.66 / 0.39       & 0.65 / 0.39  & 0.59 / 0.18 \\ 
								 & \crnn 	& 0.73 / {\bf0.70}  & 0.69 / 0.62  & 0.63 / 0.52 \\
		\hline
	\end{tabular}
	\label{tab:results_synth}
\end{table}

\vspace{-1mm}
\subsection{Synthetic dataset (\midi)}

\begin{figure}
	\centerline{
		\input{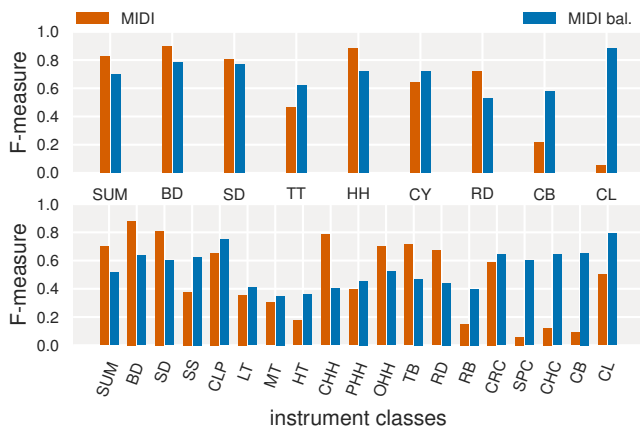}}
	\caption{Instrument class details for evaluation results on \midi\ and \midibal\ for \midlab\ and \lrglab\ instrument classes using the CRNN. First value (SUM) represents the overall sum F-measure results.}
	\label{fig:midi_midi_bal}
\end{figure}

For generating the synthetic dataset, a similar approach as in \cite{miron2013open} was employed.
Since the focus of this work is the transcription of multiple drum instruments from polyphonic music, full MIDI tracks of western popular music were used instead of MIDI drum loops.
First, every MIDI track from a freely available online collection\footnote{\texttt{http://www.midiworld.com}} was split into a drum and accompaniment track.
Using \emph{timidity++}\footnote{\texttt{http://timidity.sourceforge.net/}}, the drum tracks were rendered utilizing 57 different drum SoundFonts\footnote{\texttt{https://en.wikipedia.org/wiki/SoundFont}}.
The used SoundFonts were collected from different online sources, and great care was taken to manually check and correct the instrument mappings and overall suitability.
They cover a wide range of drum sounds from electronic drum machines (e.g. TR808), acoustic kits, and commonly used combinations.
The SoundFonts were divided into three groups for the three evaluation splits, to counter overfitting to drum kits.
The accompaniment tracks were rendered using a full General MIDI SoundFont.
Using the MIDI tracks, drum annotations as well as beat and downbeat annotations were generated.
After removing broken MIDI files, very short (< 30s) as well as very long (> 15m) tracks, the set contains 4197 tracks with an average duration of 3m 41s and a total duration of about 259h.
As with the other datasets, we only use the mixes for evaluation, while the drum solo tracks are used as additional train-only data.

Figure~\ref{fig:labeldist} shows that the general trend of the drum instrument class distribution is similar to the smaller datasets.
This is not surprising since the music is of the same broad origin (western popular music).
Since one of the goals of creating this dataset was to achieve a more balanced distribution, some additional processing is necessary.
Due to the fact that we can easily manipulate the source MIDI drum files, we can change a certain amount of instruments for several tracks to artificially balance the classes.
We did this for the \lrglab\ classes as well as for the \midlab\ classes and generated two more synthetic datasets consisting of the same tracks, but with drum instruments changes so that the classes are balanced within their respective drum instrument class system.
This was done in a way to switch instruments which have a similar expected usage frequency within a track, while keeping musicality in mind.
Ideal candidates for this are CHH and RD: exchanging them makes sense from a musical standpoint, as well in terms of usage frequency.
On the other hand, BD and CRC  are close in expected usage frequency but switching them can be questionable from a musical standpoint, depending on the music genre.
A full list of performed switches for the balanced versions can be found on the accompanying webpage.

A downside of this approach is that the instrument switches may create artificial drum patterns which are atypical for western popular music.
This can be problematic if the recurrent parts of the used CRNN architecture start to learn structures of typical drum patterns.
Since these effects are difficult to measure and in order to be able to build a large, balanced dataset, this consequence was considered acceptable.

\vspace{-2mm}
\section{Experiments}\label{sec:experiments}

\begin{table}
	\caption{F-measure results (\emph{mean}/\emph{sum}) for the CRNN model on public datasets when trained on different dataset combinations. The top part shows results for the \midlab\ class scenario, while the bottom part shows results for the \lrglab\ class scenario. Whenever the \midi\ set is mixed with real world datasets, only the 1\% subset is used, to keep a balance between different data types.}
	\centering
	\begin{tabular}{| l | c | c | c | }
		\hline
		\multicolumn{4}{| c |}{\midlab\ instrument classes} \\
		train set  & \enst & \mdb & \rbma \\
		\hline
		\all 			   & 0.61 / 0.64       & 0.68 / 0.64	  & 0.57 / 0.52 \\
		\midi 		       & 0.65 / 0.68 	   & 0.70 / 0.61      & 0.57 / 0.51 \\
		\midibal 	       & 0.61 / 0.57       & 0.66 / 0.52      & 0.56 / 0.47 \\
		\all +\midi        & 0.58 / 0.62       & 0.67 / 0.57      & 0.57 / 0.52 \\
		\all +\midibal     & 0.61 / 0.64       & 0.68 / 0.56      & 0.56 / 0.51 \\
        \ptmidi            & 0.64 / {\bf 0.69} & 0.72 / {\bf 0.68}& 0.58 / {\bf 0.56} \\
		\ptmidibal         & 0.61 / 0.63       & 0.72 / 0.67      & 0.58 / {\bf 0.56} \\
		\hline
		\hline
		\multicolumn{4}{| c |}{\lrglab\ instrument classes} \\
		train set  & \enst & \mdb & \rbma \\
		\hline
		\all 			  & 0.71 / 0.58      & 0.77 / 0.55		& 0.63 / 0.41 \\
		\midi 		      & 0.73 / 0.61      & 0.77 / 0.53      & 0.64 / 0.39 \\
		\midibal 	      & 0.70 / 0.52      & 0.76 / 0.45      & 0.63 / 0.35 \\
		\all +\midi       & 0.73 / 0.62 	 & 0.77 / 0.54      & 0.64 / 0.41 \\
		\all +\midibal    & 0.72 / 0.57      & 0.76 / 0.47      & 0.64 / 0.37 \\
        \ptmidi           & 0.74 / {\bf 0.67}& 0.78 / {\bf 0.60}& 0.64 / {\bf 0.47} \\
		\ptmidibal        & 0.74 / 0.65      & 0.78 / 0.58      & 0.64 / 0.45 \\
		\hline
	\end{tabular}
	\label{tab:results_pretrain}
\end{table}

\begin{figure*}
	\centering
	\enst\\
    \input{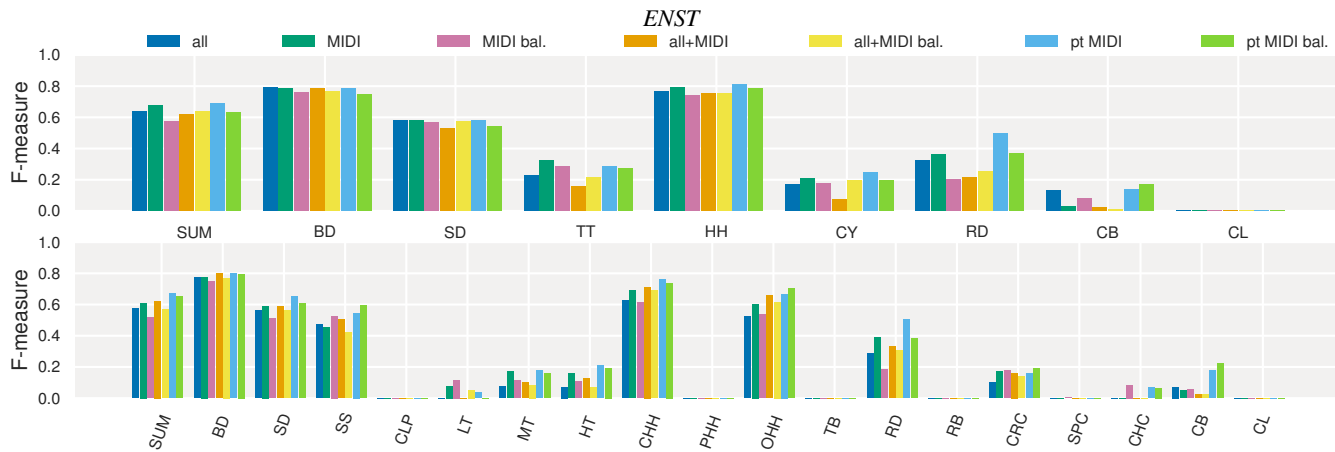}\\
	\caption[LoF entry]{This figure shows F-measure results for each instrument, for both the \midlab\ class (top) as well as the \lrglab\ class (bottom) scenarios, exemplary for the \enst\ dataset. 
		Figures for other sets are found on the accompanying webpage  (see sec.~\ref{sec:conclusion}).
		The color of bars indicates the dataset or combinations trained on: \emph{all}---three public datasets; \midi---synthetic dataset; \midibal---synthetic set with balanced classes; \emph{all+MIDI}---three public datasets plus 1\% split of synthetic dataset; \emph{all+MIDI bal.}---three public datasets plus the 1\% split of the balanced synthetic dataset; \emph{pt MIDI} and \emph{pt MIDI bal.}---pre-trained on the \midi\ and \midibal\ datasets respectively and fine tuned on \emph{all}. The first set of bars on the left (SUM) shows the overall \emph{sum} F-measure value.
	}
	\label{fig:inst_eval_cross}
\end{figure*}

The first set of experiments evaluates the implemented ADT methods on the available public datasets, using the classic three drum instrument class labels, as well as the two new drum classification schemas with \midlab\ and \lrglab\ classes, as a baseline.
As evaluation measure primarily the F-measure of the individual drum instrument onsets is used.
To calculate the overall F-measure over all instruments and all tracks of a dataset, two methods are used:
First, the mean over all instruments' F-measure (=F-measure of track), as well as the mean over all tracks' F-measure is calculated (\emph{mean}). 
Second, all false positives, false negatives, and true positives for all instruments and tracks are used to calculate a global F-measure (\emph{sum}).
These two values give insight into different aspects.
While the \emph{mean} value is more conservative for only slightly imbalanced classes, it is problematic when applied to sets containing only sparsely populated classes.
In this case, some tracks may have zero occurrences of an instrument, thus resulting in a F-measure of 1.0 when no instrument is detected by the ADT system. 
In that case, the overall \emph{mean} F-measure value for this instrument is close to 1.0 if it only occurs in a small fraction of tracks and the system never predicts it.
On the other hand, the \emph{sum} value will give a F-measure close to zero if the system never predicts an instrument, even for sparse classes---which is more desirable in this context.

The second set of experiments evaluates the performance of the ADT methods on the synthetic datasets, as well as a 1\% subset for each of the instrument classification schemas.
This will give insight in how the systems perform on the synthetic dataset and how relevant the data volume is for each of the schemas.

In the final set of experiments, models trained with different combinations of synthetic and real data will be evaluated.
The evaluation will show how well models trained on synthetic data can generalize on real world data. 
Mixing the real world datasets with the symbolic data is a first, simple approach of leveraging a balanced dataset to improve detection performance of underrepresented drum instrument classes in currently available datasets.
To be able to compare the results, models are trained on all of the public datasets (\all), the full synthetic dataset (\midi), the balanced versions of the synthetic dataset (\midibal), a mix of the public datasets and the 1\% subset of the synthetic dataset (\all+\midi), and a mix of the public datasets and a 1\% subset of the balanced synthetic datasets (\all+\midibal).
Additionally, models pre-trained on the \midi\ and \midibal\ datasets with additional refinement on the \all\ dataset were included.
We only compare a mix of the smaller public datasets to the other sets, since models trained on only one small dataset have the tendency to overfit, and thus generalize not well---which makes comparison problematic.

\vspace{-2mm}
\section{Results and Discussion}\label{sec:results}

The results of the first set of experiments is visualized in Table~\ref{tab:results_classic}, which shows the 3-fold cross-validation results for models trained on public datasets with 3, 8, and 18 labels.
The resulting F-measure values are not surprising: for the 3-class scenario the values are close to the reported values in the related work. Differences are due to slightly different models and hyper-parameter settings for training.
As expected, especially the \emph{sum} values drop for the cases of 8 and 18 classes. 
It can be observed, that the CRNN performs best for all sets in 18 class scenario and for two out of three sets for the eight class scenario.

Table~\ref{tab:results_synth} shows the results for models trained on synthetic data\-sets with 3, 8, and 18 labels. 
As expected, there is a tendency for the models trained on the 1\% subset to perform worse, especially for the CRNN.
However, this effect is not as severe as suspected.
This might be due to the fact that, while different drum kits were used, the synthetic set is still quite uniform, given its size.
The overall results for the balanced sets are worse than for the normal set. 
This is expected, since the difficulty of the balanced sets is much greater than for the imbalanced one (sparse classes can be ignored by the models without much penalty).
Figure~\ref{fig:midi_midi_bal} shows a comparison of F-measure values for individual instruments classes when training on \midi\ and \midibal\ sets.
The plot shows, that performance for underrepresented classes improves for the balanced set, which was the goal of balancing the set. 
A downside is that the performance for classes which have a higher frequency of occurrence in the \midi\ dataset decreases in most cases, which contributes to the overall decrease.
However, this effect is less severe in the \midlab\ class case.

A general trend which can be observed, especially in the scenarios with more instrument class labels, is that CRNNs consistently outperform CNNs.
Since this is true for all other experiments as well, and for reasons of clarity, we will limit the results for the next plots and tables to those of the CRNN model.

\begin{figure}[tb]
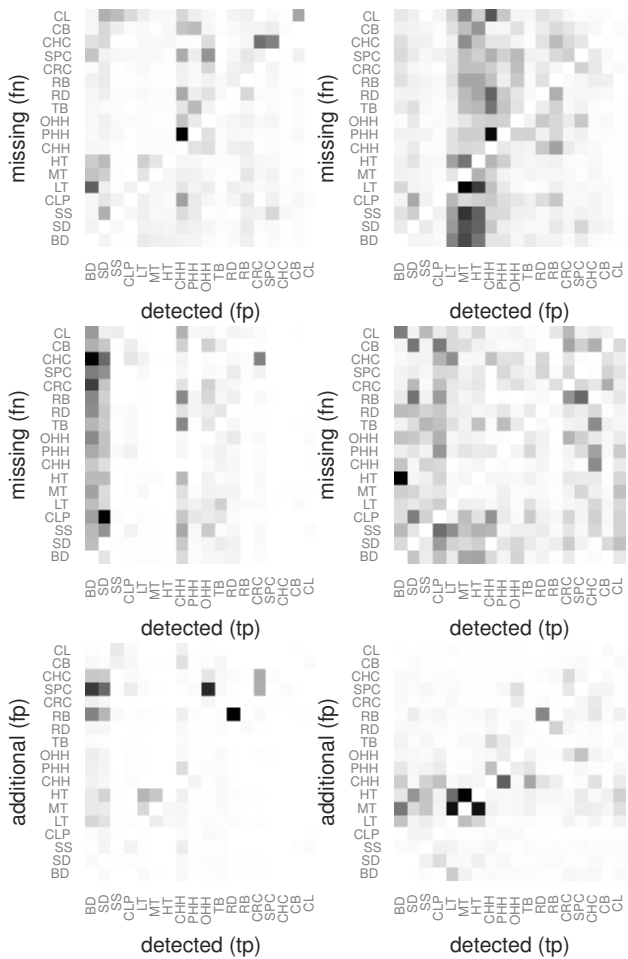

	\centering
	\input{cm-midi-conf-sml.pgf}\input{cm-midibal-conf-sml.pgf}\\
	\input{cm-midi-hid-sml.pgf}\input{cm-midibal-hid-sml.pgf}\\
	\input{cm-midi-act-sml.pgf}\input{cm-midibal-act-sml.pgf}
	\caption{Left column shows matrices for \midi\ set, right column shows matrices for \midibal\ set, both for the \lrglab\ classes scenario.
		From top to bottom, the matrices display: classic confusions (fn/fp), masking by true positives (fn/tp), and positive masking (excitement---fp/tp).}
	\label{fig:confusions}
\end{figure}

Table~\ref{tab:results_pretrain} shows the F-measure results for the CRNN model trained on different dataset combinations and evaluated on public datasets.
In figure~\ref{fig:inst_eval_cross}, a detailed look in the context of cross-datasets evaluation on instrument class basis for the \enst\ dataset is provided.
As mentioned in section~\ref{sec:experiments}, results for models trained on only one public dataset are not included in this chart. 
While the performance for those is higher, they are slightly overfitted to the individual datasets and do not generalize well to other datasets, therefore a comparison would not be meaningful.
Although an overall big performance improvement for previously underrepresented classes can not be observed, several interesting things are visible: 
\emph{(i.)}~both the models trained solely on the \midi\ and the \midibal\ datasets generalize surprisingly well to the real world dataset; 
\emph{(ii.)}~in some cases, performance improvements for underrepresented classes can be observed (e.g. for \lrglab\ classes: LT, MT, RD, CRC, CHC), when using the synthetic data; 
\emph{(iii.)}~balancing the instruments, while effective within the evaluation for the synthetic dataset, seems not to have a positive effect in the cross-dataset scenario and when mixing dataset;
and \emph{(iv.)}~using pre-training on the \midi\ set with refinement on the \all\ set, seems to produce models which are better suited to detect underrepresented classes while still performing well on other classes.

To gain more insight into which errors the systems make when classifying within the \midlab\ and \lrglab\ class systems, three sets of pseudo confusion matrices were created.
We term them \emph{pseudo} confusion matrices because one onset instance can have multiple classes, which is usually not the case for classification problems.
These three pseudo confusion matrices indicate how often
\emph{(i.)}~a false positive for another instrument was found for false negatives (classic confusions); 
\emph{(ii.)}~a true positive for another instrument was found for false negatives (onset masked or hidden);
and \emph{(iii)}~a true positive for another instrument was found for a false positive (positive masking or excitement).
Figure~\ref{fig:confusions} shows examples of these matrices for the \midi\ and \midibal\ sets in the \lrglab\ class scenario.
The images lead to intuitive conclusions: similar sounding instruments may get confused (BD/LT, CHH/PHH), instruments with energy over a wide frequency range mask more delicate instruments as well as similar sounds (HT/BD, CLP/SD), and similar sounding instruments lead to false positives (LT/MT/HT, RB/RD). 
Many of these errors may very well be made by human transcribers as well.
This also strengthens the assumption that instrument mappings are not well defined:
boundaries of the frequency range between bass drum, low, mid and high toms are not well defined, the distinction between certain cymbals is sometimes difficult even for humans, and different hi-hat sounds are sometimes only distinguishable given more context, like genre or long term relations within the piece.

To further improve performance, an ensemble of models trained on different datasets (synthetic and real, including balanced variants) can be used. 
However, experience shows that while these systems often perform best in real world scenarios and in competitions (e.g. MIREX), they give not so much insight in an evaluation scenario. 
\vspace{-2mm}
\section{Conclusion}\label{sec:conclusion}
In this work we discussed a shortcoming of current state-of-the art automatic drum transcription systems: the limitation to three drum instruments.
While this choice makes sense in the context of currently available datasets, some real world applications require transcription of more instrument classes.
To approach this shortcoming, we introduced a new and publicly available large scale synthetic dataset with balanced instrument distribution and showed that models trained on this dataset generalize well to real world data.
We further showed that balancing can improve performance for usually underrepresented classes in certain cases, while overall performance may decline.
An analysis of mistakes made by such systems was provided and further steps into this directions were discussed.
The dataset, trained models and further material are available on the accompanying webpage.\footnote{\texttt{http://ifs.tuwien.ac.at/\textasciitilde vogl/dafx2018}}

\section{Acknowledgements}\label{sec:ack}
This work has been partly funded by the Austrian FFG under the BRIDGE 1 project \emph{SmarterJam} (858514), as well as by the European Research Council (ERC) under the European Union's Horizon 2020 research and innovation programme (ERC Grant Agreement No. 670035, project \emph{CON ESPRESSIONE}).

\vspace{-1mm}

\begin{thebibliography}{10}
	
	\bibitem{VoglDWK17_DrumTransCRNN_ISMIR}
	Richard Vogl, Matthias Dorfer, Gerhard Widmer, and Peter Knees,
	\newblock ``Drum transcription via joint beat and drum modeling using
	convolutional recurrent neural networks,''
	\newblock in {\em Proc. of the 18th Intl. Soc. for Music Information Retrieval
		Conf.}, Suzhou, China, Oct. 2017.
	
	\bibitem{southall2016automatic}
	Carl Southall, Ryan Stables, and Jason Hockman,
	\newblock ``Automatic drum transcription using bidirectional recurrent neural
	networks,''
	\newblock in {\em Proc. of the 17th Intl. Soc. for Music Information Retrieval
		Conf.}, New York, NY, USA, Aug. 2016.
	
	\bibitem{SouthallSH17_DrumTransCNN_ISMIR}
	Carl Southall, Ryan Stables, and Jason Hockman,
	\newblock ``Automatic drum transcription for polyphonic recordings using soft
	attention mechanisms and convolutional neural networks,''
	\newblock in {\em Proc. of the 18th Intl. Soc. for Music Information Retrieval
		Conf.}, Suzhou, China, Oct. 2017.
	
	\bibitem{caruana1998multitask}
	Rich Caruana,
	\newblock ``Multitask learning,''
	\newblock in {\em Learning to learn}, pp. 95--133. Springer, 1998.
	
	\bibitem{gillet2004automatic}
	Olivier Gillet and Ga{\"e}l Richard,
	\newblock ``Automatic transcription of drum loops,''
	\newblock in {\em Proc. of the 29th IEEE Intl. Conf. on Acoustics, Speech, and
		Signal Processing}, Montreal, QC, Canada, May 2004.
	
	\bibitem{miron2013open}
	Marius Miron, Matthew~EP Davies, and Fabien Gouyon,
	\newblock ``An open-source drum transcription system for pure data and max
	msp,''
	\newblock in {\em Proc. of the 38th IEEE Intl. Conf. on Acoustics, Speech and
		Signal Processing}, Vancouver, BC, Canada, May 2013.
	
	\bibitem{yoshii2007drum}
	Kazuyoshi Yoshii, Masataka Goto, and Hiroshi~G Okuno,
	\newblock ``Drum sound recognition for polyphonic audio signals by adaptation
	and matching of spectrogram templates with harmonic structure suppression,''
	\newblock {\em IEEE Trans. on Audio, Speech, and Language Processing}, vol. 15,
	no. 1, pp. 333--345, 2007.
	
	\bibitem{paulus2009drum}
	Jouni Paulus and Anssi Klapuri,
	\newblock ``Drum sound detection in polyphonic music with hidden markov
	models,''
	\newblock {\em EURASIP Journal on Audio, Speech, and Music Processing}, 2009.
	
	\bibitem{Wu_2017_drumTransUnlabeledData_ismir}
	Chih-Wei Wu and Alexander Lerch,
	\newblock ``Automatic drum transcription using the student-teacher learning
	paradigm with unlabeled music data,''
	\newblock in {\em Proc. of the 18th Intl. Soc. for Music Information Retrieval
		Conf.}, Suzhou, China, Oct. 2017.
	
	\bibitem{tzanetakis2005subband}
	George Tzanetakis, Ajay Kapur, and Richard~I McWalter,
	\newblock ``Subband-based drum transcription for audio signals,''
	\newblock in {\em Proc. of the 7th IEEE Workshop on Multimedia Signal
		Processing}, Shanghai, China, Oct. 2005.
	
	\bibitem{KaliakatsosFVK12_DrumTrans_AudioMostly}
	Maximos~A. Kaliakatsos-Papakostas, Andreas Floros, Michael~N. Vrahatis, and
	Nikolaos Kanellopoulos,
	\newblock ``Real-time drums transcription with characteristic bandpass
	filtering,''
	\newblock in {\em Proc. Audio Mostly: A Conf. on Interaction with Sound},
	Corfu, Greece, 2012.
	
	\bibitem{gillet2007supervised}
	Olivier Gillet and Ga{\"e}l Richard,
	\newblock ``Supervised and unsupervised sequence modelling for drum
	transcription,''
	\newblock in {\em Proc. of the 8th Intl. Conf. on Music Information Retrieval}, Vienna, Austria, Sept. 2007.
	
	\bibitem{FitzgeraldLC_DrumTransISA_DAFX}
	Derry FitzGerald, Bob Lawlor, and Eugene Coyle,
	\newblock ``Sub-band independent subspace analysis for drum transcription,''
	\newblock in {\em Proc. Intl. Conf. on Digital Audio Effects},
	Hamburg, Germany, 2002.
	
	\bibitem{SpichZST10_DrumTransPSA_DAFX}
	Andrio Spich, Massimiliano Zanoni, Augusto Sarti, and Stefano Tubaro,
	\newblock ``Drum music transcription using prior subspace analysis and pattern
	recognition,''
	\newblock in {\em Proc. Intl. Conf. on Digital Audio Effects}, Graz,
	Austria, 2010.
	
	\bibitem{dittmar2004further}
	Christian Dittmar and Christian Uhle,
	\newblock ``Further steps towards drum transcription of polyphonic music,''
	\newblock in {\em Proc. of the 116th Audio Engineering Soc. Conv.}, Berlin,
	Germany, May 2004.
	
	\bibitem{wu2015drum}
	Chih-Wei Wu and Alexander Lerch,
	\newblock ``Drum transcription using partially fixed non-negative matrix
	factorization with template adaptation,''
	\newblock in {\em Proc. of the 16th Intl. Soc. for Music Information Retrieval
		Conf.}, M{\'a}laga, Spain, Oct. 2015.
	
	\bibitem{dittmar2014real}
	Christian Dittmar and Daniel G{\"a}rtner,
	\newblock ``Real-time transcription and separation of drum recordings based on
	nmf decomposition,''
	\newblock in {\em Proc. of the 17th Intl. Conf. on Digital Audio Effects}, Erlangen, Germany, Sept. 2014.
	
	\bibitem{wu2018survey}
	Chih-Wei Wu, Christian Dittmar, Carl Southall, Richard Vogl, Gerhard Widmer,
	Jason Hockman, Meinhard M{\"u}ller, and Alexander Lerch,
	\newblock ``An overview of automatic drum transcription,''
	\newblock {\em IEEE Trans. on Audio, Speech, and Language Processing}, vol. 26, no. 9, pp. 1457--1483, 2018.

	\bibitem{vogl2016recurrent}
	Richard Vogl, Matthias Dorfer, and Peter Knees,
	\newblock ``Recurrent neural networks for drum transcription,''
	\newblock in {\em Proc. of the 17th Intl. Soc. for Music Information Retrieval
		Conf.}, New York, NY, USA, Aug. 2016.
	
	\bibitem{vogl2017transcription}
	Richard Vogl, Matthias Dorfer, and Peter Knees,
	\newblock ``Drum transcription from polyphonic music with recurrent neural
	networks,''
	\newblock in {\em Proc. of the 42nd IEEE Intl. Conf. on Acoustics, Speech and
		Signal Processing}, New Orleans, LA, USA, Mar. 2017.
	
	\bibitem{bock2016joint}
	Sebastian B{\"o}ck, Florian Krebs, and Gerhard Widmer,
	\newblock ``Joint beat and downbeat tracking with recurrent neural networks,''
	\newblock in {\em Proc. of the 17th Intl. Soc. for Music Information Retrieval
		Conf.}, New York, NY, USA, 2016.
	
	\bibitem{WuLerch16_DrumTechniqueNMF_ISMIR}
	Chih{-}Wei Wu and Alexander Lerch,
	\newblock ``On drum playing technique detection in polyphonic mixtures,''
	\newblock in {\em Proc. of the 17th Intl. Soc. for Music Information Retrieval
		Conf.}, New York City, United States, August 2016.
	
	\bibitem{salamon2017analysis}
	Justin Salamon, Rachel~M Bittner, Jordi Bonada, Juan Jos{\'e}~Bosch Vicente,
	Emilia~G{\'o}mez Guti{\'e}rrez, and Juan~Pablo Bello,
	\newblock ``An analysis/synthesis framework for automatic f0 annotation of
	multitrack datasets,''
	\newblock in {\em Proc. of the 18th Intl. Soc. for Music Information Retrieval
		Conf.}, Suzhou, China, Oct. 2017.
	
	\bibitem{cho2014gru}
	Kyunghyun Cho, Bart van Merri{\"e}nboer, Dzmitry Bahdanau, and Yoshua Bengio,
	\newblock ``Learning phrase representations using rnn encoder–decoder for
	statistical machine translation,''
	\newblock in {\em Proc. of the Conf. on Empirical Methods in Natural Language
		Processing}, Doha, Qatar, Oct. 2014.
	
	\bibitem{hochreiter1997lstm}
	Sepp Hochreiter and J{\"u}rgen Schmidhuber,
	\newblock ``Long short-term memory,''
	\newblock {\em Neural Computation}, vol. 9, no. 8, pp. 1735--1780, Nov. 1997.
	
	\bibitem{boeck2013maximum}
	Sebastian B{\"o}ck and Gerhard Widmer,
	\newblock ``Maximum filter vibrato suppression for onset detection,''
	\newblock in {\em Proc 16th Intl Conf on Digital Audio Effects}, Maynooth,
	Ireland, Sept. 2013.
	
	\bibitem{kingma2014adam}
	Diederik~P. Kingma and Jimmy Ba,
	\newblock ``Adam: A method for stochastic optimization,''
	\newblock {\em arXiv preprint arXiv:1412.6980}, 2014.
	
	\bibitem{gillet2006enst}
	Olivier Gillet and Ga{\"e}l Richard,
	\newblock ``ENST-drums: an extensive audio-visual database for drum signals
	processing,''
	\newblock in {\em Proc. of the 7th Intl. Conf. on Music Information Retrieval}, Victoria, BC, Canada, Oct. 2006.
	
	\bibitem{southall2017mdb}
	Carl Southall, Chih-Wei Wu, Alexander Lerch, and Jason Hockman,
	\newblock ``MDB drums – an annotated subset of medleydb for automatic drum
	transcription,''
	\newblock in {\em Late Breaking/Demos, 18th Intl. Soc. for Music Information
		Retrieval Conf.}, Suzhou, China, Oct. 2017.
	
	\bibitem{bittner2014medleydb}
	Rachel~M Bittner, Justin Salamon, Mike Tierney, Matthias Mauch, Chris Cannam,
	and Juan~Pablo Bello,
	\newblock ``MedleyDB: A multitrack dataset for annotation-intensive mir
	research.,''
	\newblock in {\em Proc. of the 15th Intl. Soc. for Music Information Retrieval
		Conf.}, Taipei, Taiwan, Oct. 2014, vol.~14.
	
\end{thebibliography}

\end{document}